\documentclass[a4paper,11pt]{article}
\usepackage{setspace}
\usepackage[left = 2cm, right=2cm, top=3cm, bottom=3cm]{geometry}
\usepackage[pdftex]{graphicx}
\usepackage{cite}
\usepackage{bm}
\usepackage{float}
\usepackage{amsmath,amssymb,latexsym}
\usepackage{color}
\usepackage[T1]{fontenc}
\usepackage{wasysym}
\usepackage{physics}
\usepackage{setspace}
\usepackage{amsmath}
\usepackage{microtype}
\usepackage{relsize}
\usepackage{tikz}
\usepackage{lineno, blindtext}
\usepackage{authblk}
\usepackage[symbol]{footmisc}
\usepackage{lineno}
\usepackage{url}

\begin{document}
	
	
	
	\renewcommand{\figurename}{Fig.}
	
	\title{\color{blue}\textbf{Interfacial instabilities in confined displacements involving non-Newtonian fluids}}
	
	\author[1, $\dagger$]{Vaibhav Raj Singh Parmar}
 \affil[1]{\textit{Soft Condensed Matter Group, Raman Research Institute, C. V. Raman Avenue, Sadashivanagar, Bangalore 560 080, INDIA}}
	\author[1,*]{Ranjini Bandyopadhyay}

	\footnotetext[2]{vaibhav@rri.res.in}
	\footnotetext[1]{Corresponding Author: Ranjini Bandyopadhyay; Email: ranjini@rri.res.in}
	\maketitle
	\begin{abstract}
	The growth of interfacial instabilities during fluid displacements can be driven by gradients in pressure, viscosity and surface tension, and by applying external fields. Since displacements of non-Newtonian fluids such as polymer solutions, colloidal and granular slurries are ubiquitous in natural and industrial processes, understanding the growth mechanisms and fully-developed morphologies of interfacial patterns involving non-Newtonian fluids is extremely important. In this perspective, we focus on {displacement experiments} wherein competition{s} between capillary, viscous, elastic and frictional forces drive the onset and growth of interfacial instabilities in confined geometries. We conclude by highlighting several exciting open problems in this research area.
	\end{abstract}
		
\section{Introduction}
Ideal solids like metals follow Hooke's law, while ideal fluids like air, water and small-molecule oils satisfy Newton's law of viscosity. In contrast, non-Newtonian fluids such as blood, saliva, lava, colloidal clay suspensions, polymers and food are composed of weakly-interacting macromolecules and can simultaneously exhibit shear rate dependent viscous and elastic responses~\cite{colloidalbook}. This class of materials, ubiquitous in {our day-to-day lives and in }industrial, therapeutic and geophysical processes, {is} easily deformed by thermal stresses and often {referred to} as `soft {matter}'. 

\begin{figure}[ht]
    \centering
    \includegraphics[width=0.99\textwidth]{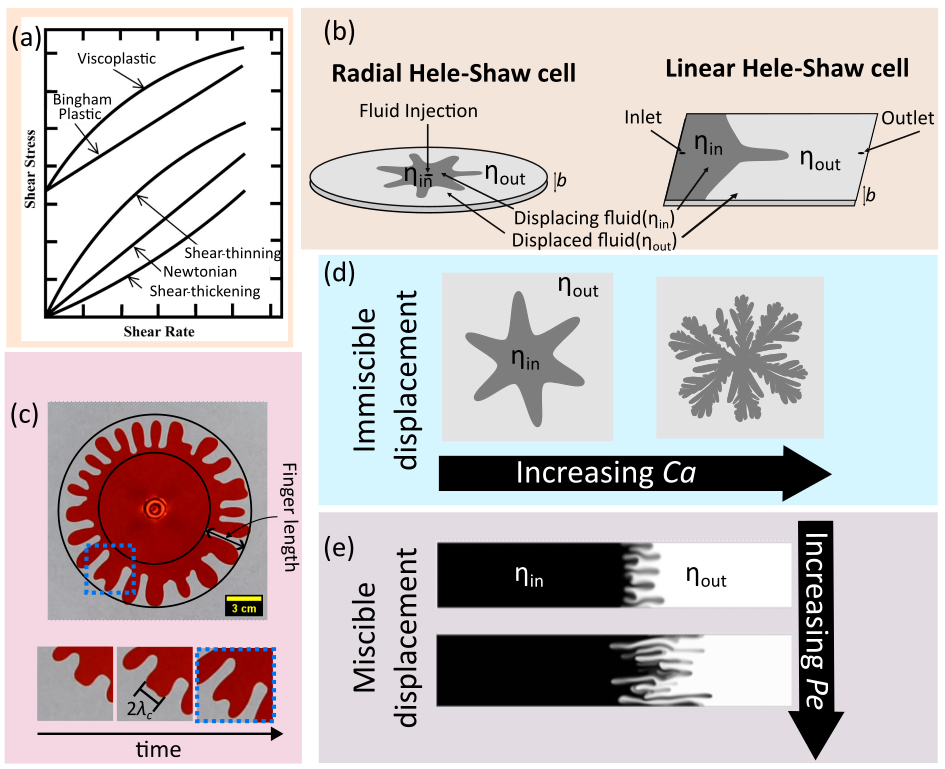}
    \caption{\textbf{Non-Newtonian flows, experimental setups and the viscous fingering instability.} \textbf{(a)} Classification of non-Newtonian fluids based on flow curves. \textbf{(b)} Schematic diagrams of radial and linear Hele-Shaw (HS) cells. \textbf{(c)} The length and width, $\lambda_c$, of a viscous finger. Formation of complex interfacial instabilities with increasing \textbf{(d)} capillary number $Ca$ and \textbf{(e)} Peclet number $Pe$. \textbf{(a,c,d)} are adapted from~\cite{Chhabra2010,C5SM00943J,chen_1989} and \textbf{(e)} is reproduced with permission from~\cite{deki_nagatsu_mishra_suzuki_2023}.
    }
    \label{fig:1}
\end{figure}

The particle-particle and particle-solvent interactions in colloidal suspensions~\cite{colloidalbook} {can lead to the formation of fragile,} heterogeneous microstructures. {Soft materials}  have non-zero relaxation times and display shear-dependent flow and deformation properties (rheology)~\cite{macosko1994rheology}. The sample viscosity can decrease with shear due to rupture of the underlying microstructures in a shear-thinning process~\cite{https://doi.org/10.1002/jctb.280450412}. Aqueous suspensions of charged colloidal clays display time-dependent mechanical properties due to a gradual and spontaneous evolution of inter-particle interactions in a physical aging process~\cite{claybook}.
Viscoplastic materials with non-zero yield stresses $\sigma_y$~\cite{COUSSOT201431} are composed of microstructures that resist flow at small deformations, but exhibit liquid-like response at stresses exceeding $\sigma_y$. The viscoelasticity of polymeric fluids arises from entanglements between neighbouring polymers that provide an effective elastic force, thereby offering enhanced resistance to flow.
Dense granular suspensions such as sand-water slurries and concrete, comprising non-Brownian particles suspended in a viscous fluid, can also exhibit viscoplastic flow~\cite{doi:10.1146/annurev-fluid-010816-060128}. Some non-Newtonian granular materials such as cornstarch suspensions display non-monotonic flow curves (stress \textit{vs.} strain rate plots), with shear-thinning rheology preceding shear-thickening {behaviour} (increase in viscosity with shear rate due to buildup of microstructure)~\cite{doi:10.1146/annurev-fluid-010816-060128}. Notable among the models to describe the gamut of viscoelastic flows (Fig.~\ref{fig:1}(a))~\cite{macosko1994rheology} is the power law model for shear-thinning and shear-thickening materials: $\eta(\dot{\gamma}) = K\dot{\gamma}^{n-1}$~\cite{macosko1994rheology}, where $\eta(\dot{\gamma})$ is the shear rate {($\dot\gamma$) dependent} viscosity of the fluid, $K$ is the flow consistency index and $n$ is a power law index. This model describes shear-thinning and shear-thickening respectively {for} $n<1$ and $n>1$. Other models for more realistic non-Newtonian fluids~\cite{macosko1994rheology} include the Cross, Carreau-Yasuda, upper convected Maxwell (UCM) and Oldroyd-B models for shear-thinning materials and the Hershel-Bulkley model for yield-stress fluids.

Multiphase flows in porous media are important in membrane separation, sanitation systems, drying and enhanced mixing processes, and also for enhanced oil recovery, sugar refining and carbon sequestration~\cite{Balmforth2001,MultiphaseFloWBook,doi:10.1146/annurev.fl.19.010187.001415,doi:10.1146/annurev-fluid-010719-060349,PINILLA2021e07614,PhysRevFluids.5.110516}. The interfacial velocity $u$ of a Newtonian fluid with viscosity
$\eta$ in a porous medium of permeability $\kappa$ is described by Darcy's law~\cite{MultiphaseFloWBook}, $u = -\frac{\kappa}{\eta} \grad P$, where $P$ is the pressure field. An interfacial instability characterised by the emergence of viscous fingers~\cite{saffman1958penetration,chen1987radial,paterson1981radial,doi:10.1146/annurev.fl.19.010187.001415} results when a fluid of 
lower viscosity $\eta_{in}$ displaces another one of higher viscosity $\eta_{out}$ in a medium of permeability $\kappa$. Wetting-induced flows, such as invasion percolation, cooperative pore filling, and corner flows were reported in capillary-dominated displacement experiment~\cite{primkulov_pahlavan_fu_zhao_macminn_juanes_2021}. Interestingly, depending on the viscosity contrast of the fluid pair, both unstable and stable flows were seen in viscosity-dominated displacement experiments~\cite{saffman1958penetration}.
The growth of viscous fingers \textit{via} successive tip-splitting results in intricate interfacial patterns that {have been} investigated in radial and linear Hele-Shaw (HS) cell geometries (Fig. 1(b)). Flow in a HS cell,  comprising a pair of transparent plates separated by an infinitesimally 
small gap $b$~\cite{paterson1981radial}, is mathematically equivalent to that in a porous medium and can be described by Darcy's Law, with $\kappa = b^2/12$~\cite{saffman1958penetration}. 

A growing finger is completely described by its width and length  (Fig.~\ref{fig:1}(c)). {The finger width, which depends} on the competition between 
destabilising viscous forces~\cite{saffman1958penetration} and stabilising capillary forces~\cite{chen1987radial}{, is} characterised by 
the most unstable wavelength, $\lambda_c = \pi b \sqrt{\frac{1}{Ca}}$, where $Ca = \frac{\Delta\eta u}{\gamma}$ is the ratio of viscous {and} capillary forces,  $\Delta \eta$ is the viscosity contrast of
the two fluids and $\gamma$ is the interfacial tension~\cite{doi:10.1146/annurev.fl.19.010187.001415}. The onset 
of instability and length of the propagating finger in a radial HS cell~\cite{bischofberger2014fingering,PhysRevLett.79.5254} are set by the viscosity ratio of the fluid pair, $\eta_{in}/\eta_{out}$.  
Experimentally observed finger widths were seen to deviate from the classical prediction~\cite{saffman1958penetration} at high capillary numbers {\textit{Ca}}~\cite{PhysRevE.88.013016,10.1063/1.3184574}. In miscible displacements ($\gamma \approx 0$ and high $Ca$), there is no well-defined interface between the displacing and displaced fluid{s} due to diffusion-driven interfacial mixing~\cite{PhysRevE.91.033006}. In such cases, interfacial instabilities depend on the ratio of advective and diffusive timescales, the Peclet number $Pe=uL/D$, with $L$ being the lengthscale characterising the system and $D$ its diffusion coefficient~\cite{sharma_nand_pramanik_chen_mishra_2020,doi:10.1146/annurev.fl.19.010187.001415}. The intricate and complex interfacial patterns that form at high $Ca$ in immiscible displacements and at high $Pe$ in miscible cases are shown in Figs.~\ref{fig:1}(d) and (e) respectively. For confined displacements involving at least one non-Newtonian fluid with a relaxation time $\tau_r$, the growth and development of interfacial patterns {{are} significantly different {when compared to} experiments involving Newtonian fluids}, and additional dimensionless numbers~\cite{PhysRevFluids.7.080701} such as Deborah number, $De = \tau_r/T$, and Weissenberg number,  $Wi = 2\tau_r \dot{\gamma}$, {{determine interfacial growth}. Here,} $T$ is the characteristic time of {the} deformation process and $\dot{\gamma}= 2U/b$~\cite{DeWi} is the characteristic deformation{/shear} rate.


In this perspective, we review {the development of }interfacial instabilities {{in displacement processes involving}} non-Newtonian fluids in Hele Shaw cells. We focus on scenarios driven by {competitions between viscous, capillary, elastic and frictional forces}{, wherein the} contributions of gravity and inertia {are negligible}. We discuss recent progress in {the study of} morphologies and { growth mechanisms} {of interfacial patterns }and conclude by {highlighting {open questions}} in this research area.

\section{Role of shear-thinning on interfacial instability}
\begin{figure}[htbp]
    \centering
    \includegraphics[width=0.99\textwidth]{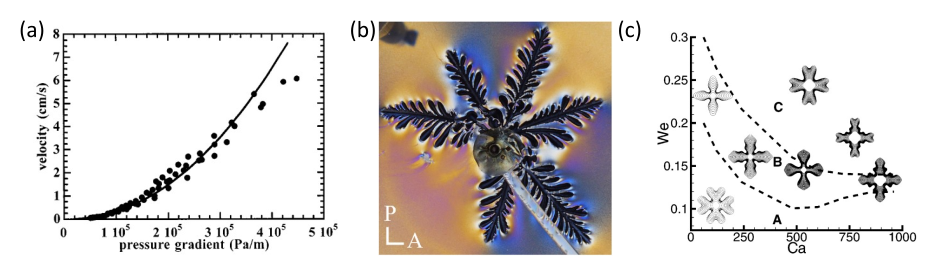}
    \caption{\textbf{Effect of shear-thinning on viscous fingering.} \textbf{(a)} Finger velocity \textit{vs.} pressure gradient during displacement of shear-thinning fluids. Strongly shear-thinning fluids deviate from the prediction of Darcy's law (solid line). \textbf{(b)} Dendritic patterns \textit{via} repeated side-branching in immiscible displacement of a shear-thinning nematic liquid crystal. \textbf{(c)} Effects of Weissenberg and capillary numbers on interfacial patterns due to immiscible displacement of a strongly shear-thinning polymeric fluid by air. Reproduced with permission from~\cite{10.1063/1.870303,doi:10.1126/sciadv.abq6820,PhysRevLett.80.1433}.
    }
    \label{fig:2}
\end{figure}

Linder $et. al.$~\cite{Lindner_2000} displaced shear-thinning solutions of the stiff polymer Xanthan by heptane oil. For low polymer concentrations, the finger width satisfies the classical prediction for two-dimensional potential flow for Newtonian fluids~\cite{mclean_saffman_1981}, but with the shear rate dependence of the viscosity, $\eta(\dot\gamma) = \eta_{eff}$, taken into account in a modified Darcy's law {(Fig.~\ref{fig:2}(a))}. Finger narrowing and deviation from classical predictions were noted with increasing polymer concentration. Due to thinning of the displaced fluid under large driving pressures, the velocity of the propagating finger is maximum at the tip. The resistance to finger growth is therefore minimum in the forward direction~\cite{PhysRevE.90.013013}, 
{which results in the} emergence of increasingly pointed fingers~\cite{BONN199560}. 
 The pressure field, which is Laplacian ($\grad^2 P = 0$) for Newtonian fluids, becomes non-Laplacian ($\grad\left[ |\grad P|^{\frac{1}{n}- 1} \grad P \right]=0$) {when} strongly shear-thinning materials {are displaced} and {is} described by a generalised Darcy's law~\cite{PhysRevE.90.013013}:
\begin{equation}
\label{eq.4}
u = - \left[ \frac{n}{(2n+1) K^{\frac{1}{n}}} \left( \frac{b}{2} \right)^{\frac{1}{n}+1} \right]{|\grad P|}^{\frac{1}{n}-1} \grad P
\end{equation}

{Simulations~\cite{PhysRevLett.80.1433,10.1063/1.1359417} of displacements of weakly shear-thinning fluids using generalised Darcy's law} showed delayed onset of tip-splitting and the formation of narrow fingers. In contrast, tip splitting was completely suppressed in strongly shear-thinning fluids.
{Finger-narrowing,} driven by shear-thinning of the displaced fluid, {competes with} finger-broadening under large driving pressures. Finger-tip oscillations, substantial side branching and dendritic pattern formation (Fig.~\ref{fig:2}(b)) {{were observed when a lyotropic chromonic liquid crystal was displaced by low-viscosity silicone oil in a radial HS cell \cite{doi:10.1126/sciadv.abq6820}}}.
 A morphological phase diagram for displacement of a strongly shear-thinning fluid is reproduced in Fig. 2(c). {Nonlinear mode coupling theory~\cite{PhysRevE.90.013013} and linear perturbation analysis~\cite{10.1063/1.2795213}} {predict the emergence} of side branches during displacement of shear thinning fluids. 
 A transition from tip-splitting to side branching growth regimes was seen {with decreasing $n$ ($0.76<n<1.0$) and increasing $De$~\cite{10.1063/1.5090772}}.  Experimental studies involving carbopol solutions, clay suspensions, liquid crystals, polymer solutions~\cite{ESLAMI201779,van1986fractal,PALAK2022100047,PALAK2023100084,PhysRevA.36.3984,doi:10.1126/sciadv.abq6820,nittmann1985fractal,10.1063/5.0102237}  and numerical simulations using Oldroyd-B fluids~\cite{10.1063/1.4977443} have reported the formation of side branches and dendrites.

{In results that are reminiscent of those reported in experiments involving a Newtonian fluid pair \cite{PhysRevLett.79.5254,bischofberger2014fingering,mehr2020mixing}, proportionate pattern growth and suppression of the fingering instability were reported when shear-thinning granular cornstarch suspensions were displaced by glycerol-water mixtures of controlled viscosity ratios, $0.06< \eta_{in}/\eta_{out} < 0.95$~\cite{PALAK2021127405}}. Interestingly, unstable to stable growth was observed when an aqueous shear-thinning Xanthan gum solution was displaced by mineral oil at $\eta_{in}/\eta_{out}>1$~\cite{10.1063/1.5133054}. 
Finite volume simulations of the displacement of a more viscous Newtonian fluid by a shear-thinning fluid showed  instability enhancement and finger coalescence~\cite{NOROUZI2018109}. 
Linear stability analysis and nonlinear simulations revealed flow stabilisation in anisotropic porous media when permeability and dispersion along the direction of flow was greater than along the transverse direction~\cite{SHOKRI20181}. 
 Interestingly, when a shear-thinning fluid was injected/withdrawn through millimetre-sized apertures at {appropriately high} flow rates, radial growth of thin equispaced protrusions (fringes) and enhanced fluid mixing were {reported}~\cite{mehr2020mixing}. {These observations {were} unaffected} by turbulence, diffusion, elongation and elastic effects. The physical principles governing {such pattern formation} remain elusive.

\section{Role of elasticity and yield stress on interfacial instability}
\begin{figure}[htbp]
    \centering
    \includegraphics[width=0.99\textwidth]{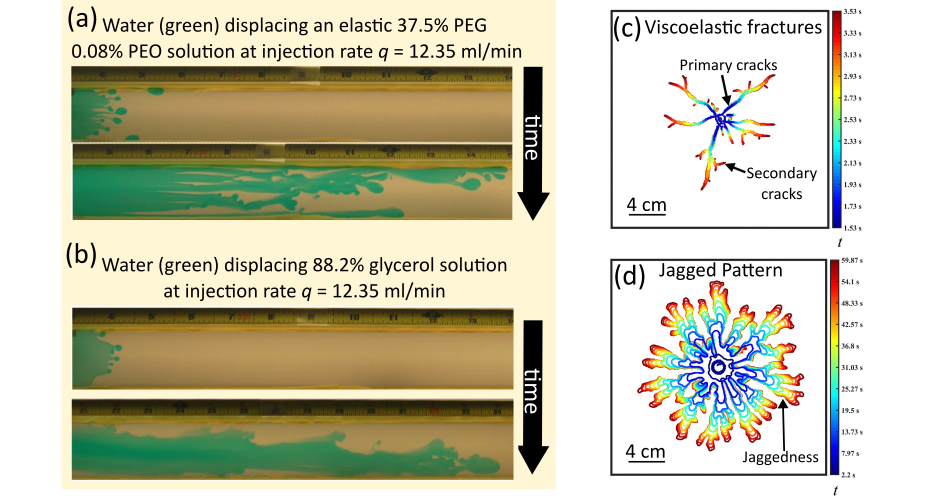}
    \caption{\textbf{Effect of elasticity on viscous fingering.} Distinct finger morphologies during displacements of \textbf{(a)} an elastic {fluid} and \textbf{(b)} a Newtonian fluid in a linear HS cell. \textbf{(c)} {Viscoelastic fractures (VEF) formed by displacing an elastic aqueous clay suspension by} miscible water. \textbf{(d)} Jagged interfacial pattern during displacement of {the} elastic clay suspension by immiscible oil. Reproduced with permission from~\cite{MALHOTRA2014125,PALAK2023100084}.
    }
    \label{fig:3}
\end{figure}

Shear-induced rupture of the underlying microscopic structures {constituting} an elastic fluid results in a reduction of sample viscosity. Distinguishing the specific roles of elasticity and shear-thinning while describing the flows of elastic fluids is therefore challenging. {The exact effect of sample elasticity on material flow can be isolated using} Boger fluids~\cite{doi:10.1146/annurev.fluid.010908.165125}, which are elastic fluids with constant viscosities comprising dilute polymers in highly viscous solvents. 
Large normal stresses are generated at the tip of a propagating finger during displacement of an elastic fluid. When compared to a Newtonian fluid {displacement experiment}, tip-splitting events occurred at lower injection pressures when a Boger fluid was displaced by immiscible air in a linear HS cell~\cite{PhysRevE.61.5439}. 
While miscible displacement of a Newtonian fluid in a linear geometry gives rise to a single dominant finger, displacement of a Boger fluid under identical conditions results in many narrow, long fingers with reduced shielding (Fig.~\ref{fig:3}(a,b))~\cite{MALHOTRA2014125}. The localised hardened zones that form at the fingertips due to very high normal stresses result in inhibited finger growth and interaction, as shown in simulations of displacement experiments involving elastic fluids with $0.005<Wi<5$~\cite{JANGIR2022103733}. Interestingly, replacing one the HS plates with an elastic membrane resulted in suppression of {viscous fingering} during displacement of a Newtonian fluid~\cite{Pihler-Puzovi2012Suppression}. In these experiments, membrane deformation in response to fluid flow generates a restoring force that oppose{s} viscous destabilisation.

Viscoelastic fluids behave as elastic solids at $De>>1$. An aqueous suspension of clay spontaneously {transitions} from a liquid to a soft solid{-like} phase with increasing concentration and due to physical aging~\cite{claybook,doi:10.1021/acs.langmuir.5b00291}. In {confined radial displacements of aqueous {clay suspensions of increasing concentrations}, a transition from viscous fingering (VF) to viscoelastic fracturing (VEF, Fig.~\ref{fig:3}(c)) was reported \cite{LEMAIRE1991}}. We note that the emergence of VEF, characterised by a small number of primary branches with perpendicular secondary offshoots, implies solid-like response of the displaced suspension. Miscible displacement of {aqueous solutions of hydrophobic polyoxyethylene}, an associating polymer solution, showed a transition from VF to VEF with increasing polymer concentration and injection pressure~\cite{PhysRevE.47.4278}. However, a VF-VEF transition was not observed in an entangled homopolymer solution, demonstrating the importance of associating networks in fracture formation. 
{During the immiscible displacement of a self-assembled transient gel, a fingering regime was reported at low injection pressures, wherein the gel was displaced gently and long range flows were generated}~\cite{C3SM51320C}. {At higher injection pressures}, the gel exhibited a solid-like response {and} was torn apart by the displacing fluid, {resulting in fracture formation} and  absence of long-range flow{s}. A dimensionless parameter $\Tilde{\lambda} = \frac{b^2(-\grad P)^{3/2}}{G\gamma^{1/2}}$~\cite{PhysRevE.81.026305} was introduced to characterise the VF-VEF transition for an upper convected Maxwell fluid~\cite{OLSSON1993125} of shear modulus $G$. While VF was observed for small values of $\Tilde{\lambda}$, the finger growth rate diverged above a critical $\Tilde{\lambda}$ due to onset of fractur{ing}.

When displaced at stresses below their yielding thresholds ${\sigma}_y$~\cite{PhysRevLett.85.314,ESLAMI201779,1999JFM...380..363C}, the solid-like/ elastic response of yield stress fluids resulted in the formation of jagged patterns and asymmetric fingers~\cite{PhysRevE.102.023105}. A ramified interface developed by the repeated tip-splitting of asymmetric fingers during the immiscible displacement of {elastic} carbopol solutions with air in a linear HS cell~\cite{PhysRevE.102.023105}. {In such displacement experiments}, the width of the finger is set by $\sqrt{\gamma b/\sigma_y}$ and flows become increasingly unstable with increasing roughness of the HS cell plate~\cite{Maleki-Jirsaraei_2005,DUFRESNE2023104970}. 
{Since the low interfacial tension in a miscible displacement experiment reduces fracture energy,} viscoelastic fractures (VEF) are more likely to form in miscible rather than in immiscible displacements {of yield stress fluids below their yielding points}~\cite{BALL2021104492}. A transition from VEF to pattern growth driven by fingertip splitting was reported when a miscible displacing fluid was injected at a rate that resulted in yielding of an emulsion~\cite{10.1063/1.1709543}. {When elastic foam  was displaced by air at increasing driving pressures,} jagged interfacial patterns (Fig.~\ref{fig:3}(d)) observed below the yielding point transitioned to smooth fingers (Fig.~\ref{fig:1}(c)) in the viscosity-dominated regime~\cite{PhysRevLett.72.3347}. 
When aqueous clay suspensions of increasing ages were radially displaced by miscible water at low injection rates, a sequence of interfacial patterns, from dense viscous to dendrites and eventually VEFs, were observed~\cite{PALAK2022100047,PALAK2023100084}. In contrast, when the aging suspensions were displaced by immiscible mineral oil, a transition from flower (Fig.~\ref{fig:1}(c)) to jagged (Fig.~\ref{fig:3}(d)) patterns was observed. 

\section{Role of shear-thickening on interfacial instability}
\begin{figure}[ht]
    \centering
    \includegraphics[width=0.99\textwidth]{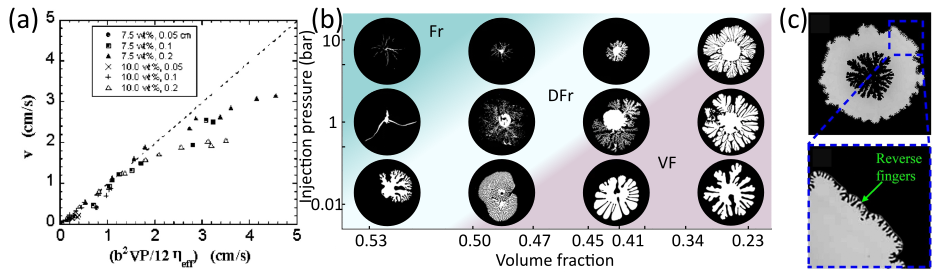}
    \caption{\textbf{Effect of shear-thickening on interfacial instability.} \textbf{(a)} Average finger velocity as a function of $b^2 \grad P / 12 \eta_{eff}$~\cite{10.1063/1.1894407}. {Strongly shear-thickening fluids show substantial deviations from Darcy's law (dashed line)}. \textbf{(b)} Phase diagram in the pressure-volume fraction plane displaying viscous fingers (VF), dendritic fractures (DFr) and large-scale fracturing (Fr). \textbf{(c)} Formation of reverse fingers {due to invasion of air} at the outermost interface between a discontinuously shear-thickening cornstarch suspension and atmospheric air. Reproduced with permission from~\cite{10.1063/1.1894407,ozturk2020flow,PALAK2023130926}.
    }
    \label{fig:4}
\end{figure}

Displacement of a shear-thickening fluid produces wide fingers that grow $via$ {rapid} tip-splitting events~\cite{PhysRevE.67.026313,PhysRevE.90.013013,doi:10.1021/bk-2004-0869.ch020}.
 For {strongly} shear-thickening fluids, finger propagation slows down substantially to a value below the prediction of the modified Darcy's law (Fig.~\ref{fig:4}(a),~\cite{10.1063/1.1709543}). 

Shear-thickening fluids display continuous shear-thickening (CST), discontinuous shear-thickening (DST), and shear-jamming (SJ) {flow regimes} with increasing particle volume fraction $\phi$ and applied shear stress $\sigma$/ rate $\dot{\gamma}$~\cite{doi:10.1146/annurev-fluid-010816-060128}. 
These distinct rheological regimes of cornstarch suspensions can be achieved in radial displacements experiments by appropriately controlling the injection pressures of the displacing Newtonian fluids~\cite{ozturk2020flow,10.1063/5.0152800,PALAK2023130926}. When cornstarch suspensions in CST, DST and SJ were displaced by immiscible air in separate experiments, three distinct interfacial patterns, \textit{viz.}, viscous fingers (VF), dendritic fracturing (DFr) and system-wide fracturing (Fr)(Fig.~\ref{fig:4}(b)) were produced~\cite{ozturk2020flow}. A transition from VF to {VEF} was reported in cornstarch suspensions when oil was used as a displacing fluid~\cite{10.1063/5.0152800}.
Miscible displacements of discontinuously shear thickening cornstarch suspensions with water~\cite{PALAK2023130926} showed transient invasion of air into the suspension and formation of reverse fingers at the outermost interface between the suspension and atmospheric air (Fig.~\ref{fig:4}(c)). Since the suspension {was} in the DST regime, the stresses {due to} injection of the displacing fluid {are expected to} propagate through force networks. Suspension dilation~\cite{doi:10.1146/annurev-fluid-010816-060128}, {followed by} a surface tension driven restoring force, drives the formation of the observed reverse fingers. To the best of our knowledge, there are no complete theoretical or numerical descriptions of dendritic fracturing and reverse finger formation in Hele-Shaw flows.

\section{Interfacial instabilities in multiphase frictional and viscous flows}
In this section, we focus on displacement processes involving movable athermal grains. Advancements in the understanding of multiphase flows in granular media have been summarised recently~\cite{PhysRevFluids.5.110516,doi:10.1146/annurev-fluid-010518-040342}. 
\begin{figure}[ht]
    \centering
    \includegraphics[width=0.49\textwidth]{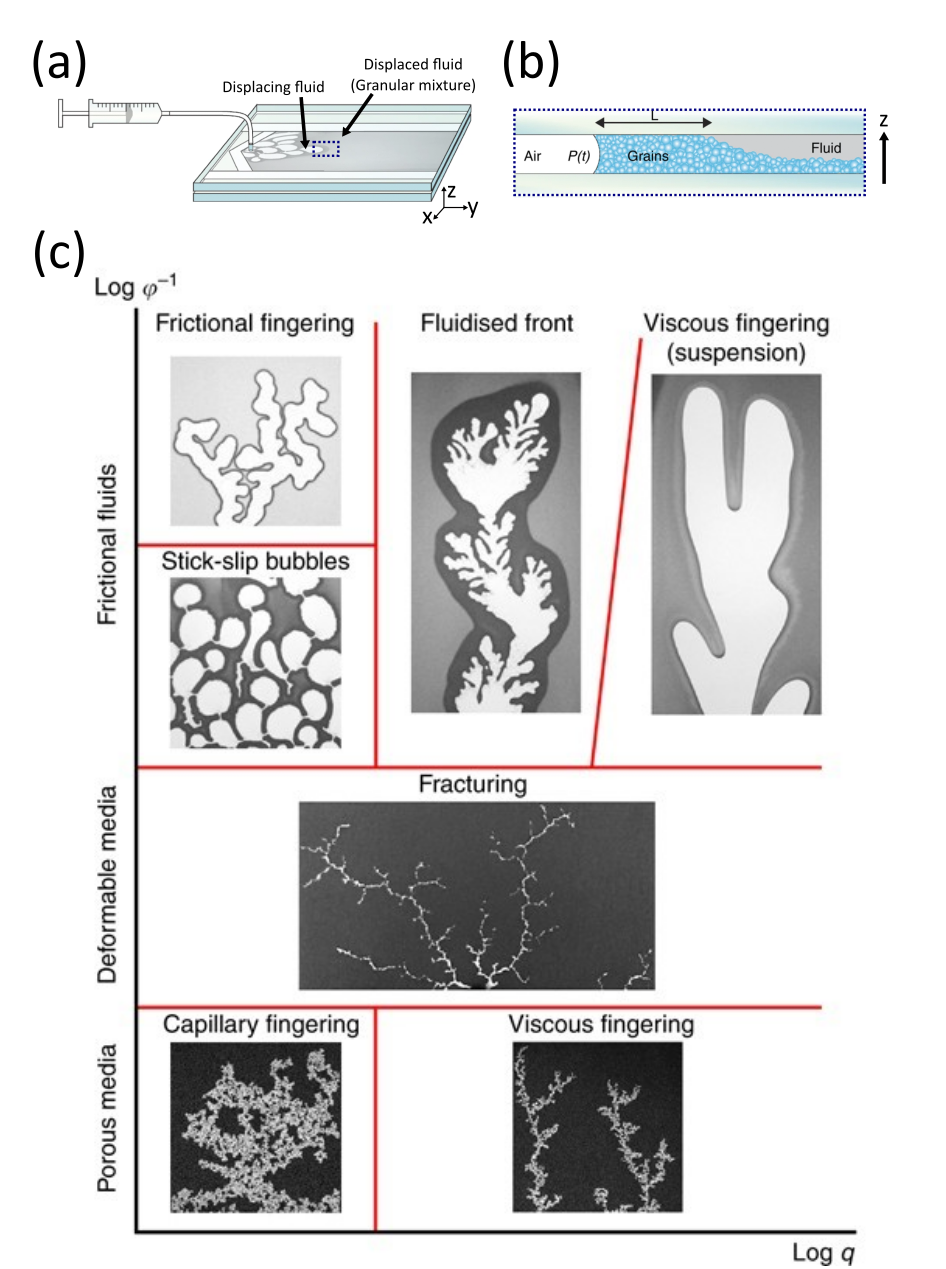}
    \caption{\textbf{Pattern formation during displacement of deformable granular matter.} {Schematic illustrations of \textbf{(a)} a rectilinear Hele-Shaw cell used for granular displacement experiments and \textbf{(b)} the formation of a compact front of grains due to the bulldozing effect.}  \textbf{(c)} Phase diagram in the inverse of volume fraction, $\phi^{-1}$, and injection rate, $q$, plane, displaying various interfacial pattern morphologies during displacement {of} deformable granular media undergoing drainage. Reproduced with permission from~\cite{sandnes2011patterns}. }
    \label{fig:5}
\end{figure}
{For granular particles that} move during displacement~\cite{PhysRevLett.99.038001,sandnes2011patterns,PhysRevFluids.5.110516,10.3389/fphy.2015.00081,PhysRevLett.108.258001}, {the inter-grain solid friction} assumes a crucial role in understanding their flow~\cite{CHEVALIER200963}. At low grain fraction $\phi$, the flow profile was reported to obey modified Darcy's law, with grain sizes determining the lengthscale of the instability~\cite{PhysRevLett.99.174501}. Polydisperse glass beads, suspended in a glycerol-water mixture {and} displaced by air, accumulated at the interface {to form} a compact front {(contact angle $\theta>90^\circ$, experimental geometry reproduced in Fig.~\ref{fig:5}(a))~\cite{sandnes2011patterns}}. This phenomenon, known as bulldozing, occurs when the threshold capillary pressure exceeds inter-grain frictional resistance to sliding and rearrangements, and {is displayed} in Fig.~\ref{fig:5}(b). Friction-dominated finger dynamics was reported at low injection rates, and a transition from a creeping fingering process to intermittent stick-slip bubble dynamics was seen with increasing $\phi$ (Fig.~\ref{fig:5}(c)). As the viscous forces become significant with increasing injection rates, fluidisation of the the compact granular front and viscous fingering were reported~\cite{sandnes2011patterns,CHEVALIER200963}. Bulldozing causes frictional instability in the flow at all injection rates, however, viscous forces can stabilise the flow if $\eta_{in} > \eta_{out}$. A study investigated the competition between stabilising viscous forces and destabilising frictional forces by injecting glycerol-water mixture into dry hydrophobic grains~\cite{zhang2023frictional}. A transition from frictional fingers to stable displacement with radial spikes of grains was reported with increasing injection pressure, $\eta_{in}$ and {volume fraction inverse, }$\phi^{-1}$. The response of the displaced {medium} changed from fluid-like to solid-like at high $\phi$, with the emergence of fractures at high injection pressures. These results are depicted in Fig.~\ref{fig:5}(c). 

When spherical glass beads were displaced by air in a radial HS cell in the zero surface tension limit ($\theta \approx 0^{\circ}$), very sharp fingers grew via successive tip splitting when the injection pressure exceeded the {suspension }yield stress~\cite{cheng2008towards}. {In contrast to Newtonian fluid displacement, the fully developed pattern took on a highly ramified structure, but {became}} smoother with wider fingers at higher injection pressures. Unlike Newtonian displacement patterns~\cite{Nand2022}, the characteristic finger width in granular fingering is independent of the HS cell gap~\cite{cheng2008towards}. The bulldozing effect and frictional fingers were absent in this study~\cite{cheng2008towards} due to the negligible threshold capillary pressure. {The observed differences between displacements of granular fluids and Newtonian fluids arise from the distinct dissipation mechanisms, frictional \textit{vs.} viscous, in the two processes.}

\section{Summary and open questions}

In this perspective, we have focussed on the distinct morphologies and growth mechanisms of interfacial patterns that result during the displacement  of non-Newtonian colloidal and granular fluids in radial and linear Hele-Shaw cells. {These studies clearly demonstrate the important role of fluid rheology in pattern formation.} There are many other experimental setups used to study fluid-fluid displacements~\cite{morrow_moroney_dallaston_mccue_2021} that we have not discussed due to space constraints. We would like to refer the reader to the literature involving lifting Hele-Shaw cells, wherein the cell gap is time-dependent~\cite{Singh_2020,PhysRevLett.124.248006} and inertial effects {influence the growth of} the emergent interfacial patterns~\cite{PhysRevFluids.2.014003}. Injecting the displacing fluid at a non-uniform rate can suppress the viscous fingering instability for both Newtonian and non-Newtonian displacements~\cite{PhysRevFluids.6.033901,10.1063/5.0124066}. Tapered Hele-Shaw cells, with  cell plates that are not parallel, have been reported to reduce the {viscous fingering} instability~\cite{Al-Housseiny2012}. A multiport lifted Hele-Shaw cell with multiple source holes has been used to engineer complex interlinked meshes~\cite{Islam2017}, while experiments with rotating {Hele-Shaw} cells have been performed to study density-driven instabilities~\cite{WOS:A1996WA83900050}. Displacement experiments involving a pair of chemically reactive fluids in a {Hele-Shaw} cell have reported dramatic changes in the morphology and growth of viscous fingers~\cite{doi:10.1146/annurev-fluid-010719-060349}. Electro-osmotic flows have been induced by applying electric fields to control {viscous fingers} in Newtonian~\cite{Gao2019} and non-Newtonian fluid displacements~\cite{LI2022105204}.
Among other cell configurations not discussed in this perspective are fluid withdrawal experiments, wherein viscous and particle-driven instabilities have been reported~\cite{PhysRevFluids.3.110502,PhysRevLett.118.074501}.

While the role of material shear rheology has been the main focus in studies of the {viscous fingering} instability, exploiting the extensional rheology of non-Newtonian fluids holds a lot of promise in our ability to control fluid displacement processes~\cite{PhysRevE.61.5439,KAZEMI2023211}. The effect of interfacial flows in the presence of surface active agents~\cite{PhysRevE.108.025104} is an area that also remains relatively less explored for non-Newtonian fluid displacements. Local events in viscous fingering~\cite{PhysRevE.61.5439,doi:10.1146/annurev.fl.19.010187.001415} are {comparatively better} studied than for viscoelastic fracture formation. 
Analytical and numerical investigations are essential for acquiring fundamental knowledge of elasticity-driven interfacial instabilities. Appropriate models need to be developed to explain other novel phenomena that have been observed in experiments, for example, the formation of fringes~\cite{mehr2020mixing}, asymmetric~\cite{PhysRevE.102.023105} and reverse fingers~\cite{PALAK2023130926}, and dendritic fractures~\cite{ozturk2020flow}. Many non-Newtonian fluids show chaotic flows at low Reynolds numbers and are known to generate anomalous flow resistance~\cite{doi:10.1126/sciadv.abj2619}. The formation of viscous fingers can expedite fluid mixing at low Reynolds numbers~\cite{PhysRevLett.106.194502}. Detailed studies on how chaotic flows affect the viscous fingering instability and fluid mixing are therefore {critical}. Tuning interactions between the macromolecular constituents of non-Newtonian fluids modifies fluid rheology~\cite{D1SM00987G} and can be exploited to pre-program interfacial instability patterns. The ability to control, enhance and suppress finger formation, and therefore fluid displacement efficiency~\cite{10.1063/1.4977443}, is {crucial} in several industrial processes, calling for further rigorous study of the mechanisms of formation of complex interfacial instability patterns. 


\section*{Data Availability}
No new data were created or analysed in this article.

\section*{Acknowledgments}	 
The authors thank Raman Research Institute, Bangalore, India, for financial support.

\bibliographystyle{elsarticle-num}
\bibliography{ref}

\end{document}